\documentclass[sigconf]{acmart}
\usepackage[ruled,linesnumbered]{algorithm2e}
\usepackage{multirow}
\usepackage{colortbl}

\renewcommand\footnotetextcopyrightpermission[1]{}

\AtBeginDocument{%
  \providecommand\BibTeX{{%
    \normalfont B\kern-0.5em{\scshape i\kern-0.25em b}\kern-0.8em\TeX}}}

\setcopyright{acmcopyright}
\copyrightyear{2022}
\acmYear{2022}
\acmDOI{XXXXXXX.XXXXXXX}

\acmConference[MM'22]{the 30th ACM International Conference on Multimedia}{October 10--14,
  2022}{Lisbon, Port}
\acmPrice{15.00}
\acmISBN{978-1-4503-XXXX-X/18/06}

\begin{document}

\title{A Comprehensive Benchmark Analysis for Sand Dust Image Reconstruction}

\author{Yazhong Si}
\orcid{0000-0002-4454-1605}
\affiliation{%
  \institution{Hebei University of Technology}
  \city{Tianjin}
  \country{China}
  \postcode{300401}
}

\author{Fan Yang}
\affiliation{%
  \institution{Hebei University of Technology}
  \city{Tianjin}
  \country{China}
  \postcode{300401}
}

\author{Ya Guo}
\affiliation{%
  \institution{Hebei University of Technology}
  \city{Tianjin}
  \country{China}
  \postcode{300401}
}

\author{Wei Zhang}
\affiliation{%
  \institution{Hebei University of Technology}
  \city{Tianjin}
  \country{China}
  \postcode{300401}
}
\author{Yipu Yang}
\affiliation{%
  \institution{Hebei University of Technology}
  \city{Tianjin}
  \country{China}
  \postcode{300401}
}

\renewcommand{\shortauthors}{Si and Yang, et al.}

\begin{abstract}
Numerous sand dust image enhancement algorithms have been proposed in recent years. To our best knowledge, however, most methods evaluated their performance with no-reference way using few selected real-world images from internet. It is unclear how to quantitatively analysis the performance of the algorithms in a supervised way and how we could gauge the progress in the field. Moreover, due to the absence of large-scale benchmark datasets, there is no well-known report of data-driven based method for single image sand dust removal up till now. To advance the development of deep learning-based algorithms for sand dust image reconstruction, while enabling supervised objective evaluation of algorithm performance. In this paper, we presented a comprehensive perceptual study and analysis of real-world sand dust images, then constructed a \emph{Sand-dust Image Reconstruction Benchmark (SIRB)} for training \emph{Convolutional Neural Networks (CNNs)} and evaluating algorithms performance. In addition, we adopted the existing image transformation neural network trained on SIRB as baseline to illustrate the generalization of SIRB for training CNNs. Finally, we conducted the qualitative and quantitative evaluation to demonstrate the performance and limitations of the current sand dust removal algorithms, which shed light on future research in the field of sand dust image reconstruction.
\end{abstract}



\begin{teaserfigure}
  \includegraphics[width=1\columnwidth]{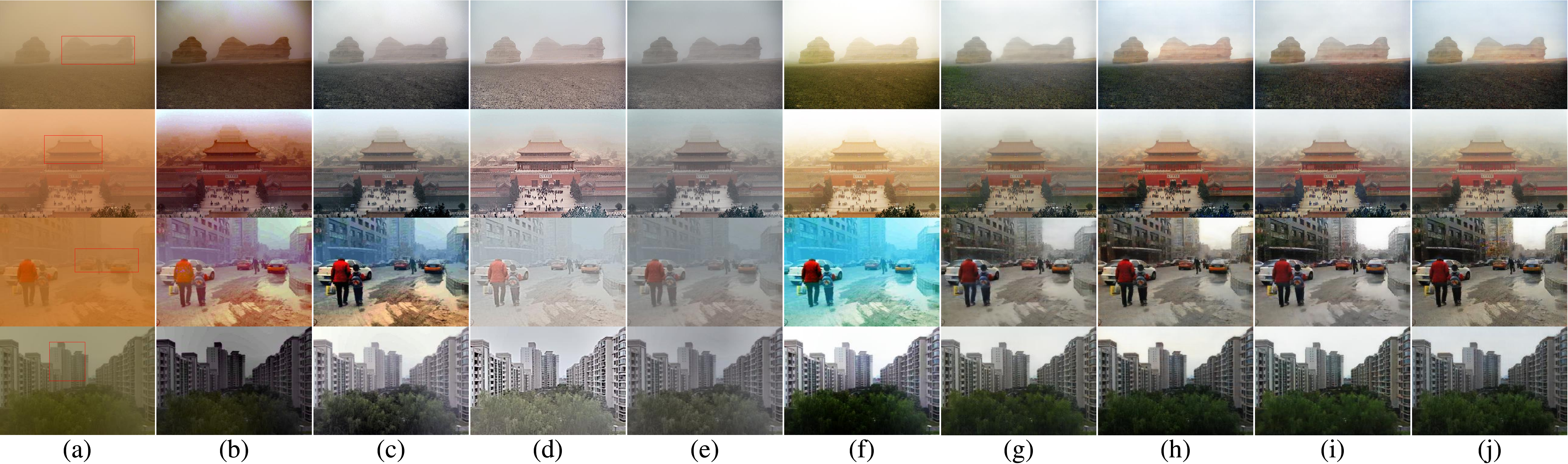}
  \caption{The comparisons on sand dust degraded images from SIRB. (a) Sand dust images; (b) CIDC \cite{ref2}; (c) FBE \cite{ref14}; (d) HRDCP \cite{ref4};  (e) NGT \cite{ref10}; (f) TTFIO \cite{ref58}; (g) Pix2pix-L \cite{ref19}; (h) Pix2pix-M \cite{ref19}; (i) Pix2pix-D \cite{ref19}; (j) Pix2pix-H \cite{ref19}.}
  \label{Fig.1}
\end{teaserfigure}

\maketitle

\section{Introduction}
In sandstorm weather, affected by Mai scattering, the blue-violet light is absorbed by sand dust particles much more quickly compared with red and orange light. \cite{ref1} The images captured in outdoor scenes often suffer from color cast, poor visibility, and fainted surfaces, which play an adverse role in various remote-based computer vision tasks and seriously interferes with the performance of intelligent information processing system. In contrast to haze imaging conditions, where the illumination is dominated by global atmospheric light with the same R, G, B values, degraded images in sand dust weather have obvious prior characteristics for shifting, concentration and sequential due to varying degrees of attenuation of R, G, B values. Eliminating color cast and hazy is essential for enhancing the visibility of sand dust images, which, however, due to its heavily ill-posed nature is very challenging.

To address the issues, many studies have been proposed \cite{ref2,ref3,ref4,ref5,ref6,ref7,ref8,ref9,ref10,ref11,ref12,ref13,ref14}, which can be roughly divided into two strategies including image restoration \cite{ref2,ref3,ref4,ref5,ref6} and image enhancement \cite{ref7,ref8,ref9,ref10,ref11,ref12,ref13,ref14}. The sand dust image restoration algorithms are designed based on the atmospheric scattering model. Commonly, they estimate the transmission and global atmosphere light value via prior knowledge \cite{ref15}, and then put the intermediate variables into the physics model to restore the clear images. The enhancement-based strategies improve the clarity of sand dust images by compensating the color channel, balancing the brightness, and stretching the contrast. These algorithms all can improve the image quality to a certain extent, however, there are still exist some issues cannot be ignored. In general, the atmospheric scattering model is adopted to describe the generation process of haze image, which is slightly far-fetched used in sand dust removal task. For image enhancement algorithms, the processed images often suffer color distortion due to the scholars are excessive focus on sandstorm removal without considering the overall visual perception. Image sand dust removal, as an ill-posed issue, has not been effectively solved.

Recent years have witnessed significant advances in deep learning, which has attracted extensive attention in the field of image processing \cite{ref16,ref17,ref18,ref19,ref20} because of its strong ability to fit data. Many scholars adopted deep learning to process degraded images in complex environments, such as image dehazing \cite{ref21,ref22,ref23,ref24,ref25}, image deraining \cite{ref26,ref27,ref28,ref29,ref30} and underwater image enhancement \cite{ref31,ref32,ref33,ref34,ref35}, etc. The image reconstruction algorithms based on deep learning in complex environment have achieved remarkable performance, which mainly benefit from the proposal of some outstanding benchmarks \cite{ref36,ref37,ref38,ref39,ref40,ref41,ref42,ref43,ref44}. Unfortunately, the development of learning based sand dust image reconstruction algorithms is lagging due to the absence of large-scale benchmark datasets. Additionally, the lack of publicly available dataset makes it is hard to evaluate the performance of sand dust removal algorithms using full-reference metrics.

To bridge the gaps, in this paper, we simulated the distribution of sand dust images, and proposed a large-scale sand dust image reconstruction benchmark for evaluating algorithms performance and training CNNs. As shown in Figure \ref{Fig.1}, we trained the CNNs on the constructed dataset, and compared the results with the prior algorithms. One can see that CNNs have obvious advantages in terms of color correction and dust removal. 

The main contributions of this work are summarized as follows:
\begin{itemize}
\item We proposed a single image sand dust removal benchmark named \emph{Sand-dust Image Reconstruction Benchmark (SIRB)} which contains 16000 synthetic images and 230 real-world sand dust images, covering a wide range of sandstorm degraded scenes;
\item We introduced a set of full-reference and non-reference strategies to evaluate the performance of the algorithms both qualitatively and quantitatively, which provide comprehensive insights into the strengths and limitations of current sand dust image enhancement methods;
\item With the constructed SIRB, we trained a classical image transformation CNNs model as baseline and compared the CNNs with the existing sand dust removal methods. The evaluation and analysis will provide some constructive inspiration for future research of data-driven based sand dust removal algorithms.
\end{itemize}

\section{Existing Methodology and Evaluation Metric}
\label{sec:headings}

\subsection{Single image sand dust removal algorithms}
Recently, numerous sand dust image methods have been proposed \cite{ref2,ref3,ref4,ref5,ref6,ref7,ref8,ref9,ref10,ref11,ref12,ref13,ref14} which can be roughly divided into two classes. One line of research is the restoration algorithms, and another line is tries to modify the pixel values using the prior knowledge to enhance visual quality of the image.

\subsubsection{Physical model based algorithms}
Image restoration algorithms are designed based on atmospheric scattering model, the physical model was proposed by McCartney \cite{ref45} is widely used in haze removal task and can be written as:
\begin{equation}
	I(x) = J(x)t(x) + A(1 - t(x))
	\label{eq:1}
\end{equation}
To obtain the haze-free image, the atmospheric scattering model can be rewritten as the following:
\begin{equation}
	J(x) = A + \frac{{I\left( x \right) - A}}{{t\left( x \right)}}
	\label{eq:2}
\end{equation}
where $I(x)$ is haze image; $J(x)$ is the corresponding haze-free image of $I(x)$; $t(x)$ is the transmission; $A$ is the global atmosphere light of the haze image, it is a scalar and can be expressed as:
\begin{equation}
    A = \left( {{A_R},{A_G},{A_B}} \right)
    \label{eq:3}
\end{equation}
when $A_R$=$A_G$=$A_B$, we can get the color balanced haze images. But in sand dust weather, the principle of $A_R$\textgreater$A_G$\textgreater$A_B$ always be followed. The main task of sand dust image restoration using atmospheric scattering model is how to estimate t(x) and A accurately.

At present, the \emph{Dark Channel Prior (DCP)} algorithm \cite{ref15} is commonly used to estimate the necessary parameters in the physical model. Based on halo-reduced dark channel prior, Shi et al. \cite{ref4} proposed a method for sand dust image enhancement, which including color correction in the LAB color space via gray world theory and dust removal using a halo-reduced DCP dehazing algorithm. Gao et al. \cite{ref5} presented a sand dust image restoration algorithm based on reversing the blue channel prior which improves the visibility of sand dust degraded images via DCP theory. Yu et al. \cite{ref6} proposed a method based on atmospheric scattering model and information loss constraint to eliminate the sand dust. Specifically, they optimized the initial atmospheric light value using the constraint of information loss and estimated the coarse transmission with DCP theory simultaneously, and then refined the coarse transmission by the fast guide filter.

However, there are some issues should be of concern. Although researchers have balanced the color of sand dust images before using the atmospheric scattering model, it is still a little far-fetched in logic to apply the dehazing model on sand dust removal task. Meanwhile, the existing prior knowledge cannot effectively estimate the transmission of the sky area, it is easily causing the color distortion of the output images in sky or highlight regions.

\subsubsection{Non-model based algorithms}
Non-model based algorithms aim to adjust the pixel values using the prior knowledge to enhance visual quality of sandstorm degraded images. Cheng et al. \cite{ref7} proposed a fast sand dust image enhancement algorithm based on channel compensation and guided image filtering, which can recover the fading characteristics of sand dust degraded images in a short time and improve the clarity of the images. Park et al. \cite{ref11} presented a method using successive color balance with coincident chromatic histogram for sand dust image enhancement. To enhance the degraded images, Xu et al. \cite{ref13} designed an alternative edge-preserving color image decomposition model based on the \emph{tensor least squares (TLS)} framework and the statistical feature of natural color images. Built upon fusion principles, Fu et al. \cite{ref14} proposed an approach by fusing the inputs with the weight maps which contain important features to increase the visual quality of the sand dust degraded image.

In contrast to image dehazing tasks \cite{ref15,ref46,ref47}, sand dust enhancement algorithms require to consider more external factors. Despite the prolific work, both the comprehensive study and insightful analysis of sand dust image enhancement algorithms remain largely unsatisfactory due to the complex distribution of sand dust degraded images. The currently algorithms of sand dust enhancement can mitigate the visual impact of sand to a certain extent, however, there still existing some issues in the algorithms for incomplete dust removal, loss of detail information and color distortion. 

\begin{figure}
	\centering
	\includegraphics[width=1\columnwidth]{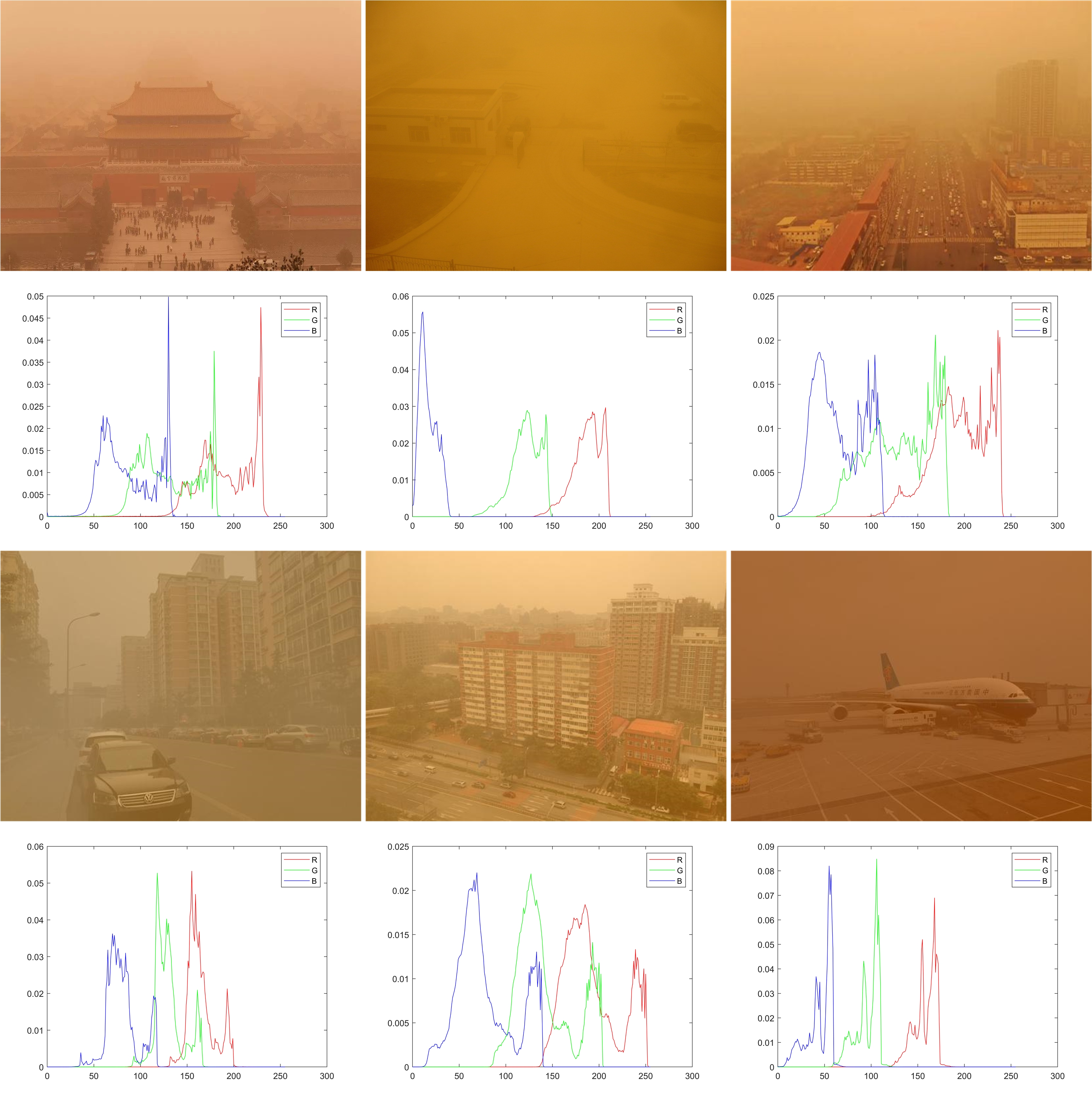}
	\caption{Comparison examples of real-world and synthetic sand dust images. The top two rows represent real sand dust images and the corresponding RGB distribution histograms; The bottom two rows are the synthetic degraded images and the corresponding histograms.}
	\label{Fig.2}	
\end{figure}

\subsection{Sand dust image quality evaluation}
\subsubsection{Non-reference metrics}
Non-reference metrics refers to directly calculate the visual quality of the results when the reference images does not exist. Due to the lack of sandstorm benchmark dataset, most researchers using only non-reference metrics to evaluate the performance of sand dust removal algorithms. In addition to the basic evaluation metrics such as \emph{Average Gradient (AG), Edge Intensity (EI)} and \emph{Information Entropy (IE)}, the non-reference evaluation strategies for \emph{Natural Image Quality Evaluator (NIQE)} \cite{ref48}, \emph{Spatial–Spectral Entropy-based Quality (SSEQ)} \cite{ref49}, and \emph{Blind Image Quality Indices (BIQI)} \cite{ref50} are also effective to measure the algorithms performance.

\subsubsection{Full-reference metrics}
Full-reference evaluation metrics often be employed to measure the performance of the image processing algorithms. \emph{Mean Square Error (MSE)} and \emph{Peak-Signal to Noise Ratio (PSNR)} are used to calculate the errors in pixel level; \emph{Structural Similarity (SSIM)} \cite{ref51} and \emph{Feature Similarity-color (FSIMc)} \cite{ref52} can measure the similarity structural between the output and reference images; And CIE94 \cite{ref53} and CIEDE2000 \cite{ref54} often be chosen to calculate the chromatic aberration. In fact, it is practically impossible to simultaneously capture a real degraded scene and the corresponding ground truth image for complex environment. That increases the difficulty for researchers to compute the full-reference metrics. For this reason, many datasets \cite{ref36,ref37,ref38,ref39,ref40,ref41,ref42,ref43,ref44} were presented to bridge the gap, which greatly promote the development of the algorithms in related fields.

\begin{figure*}
	\centering
	\includegraphics[width=2\columnwidth]{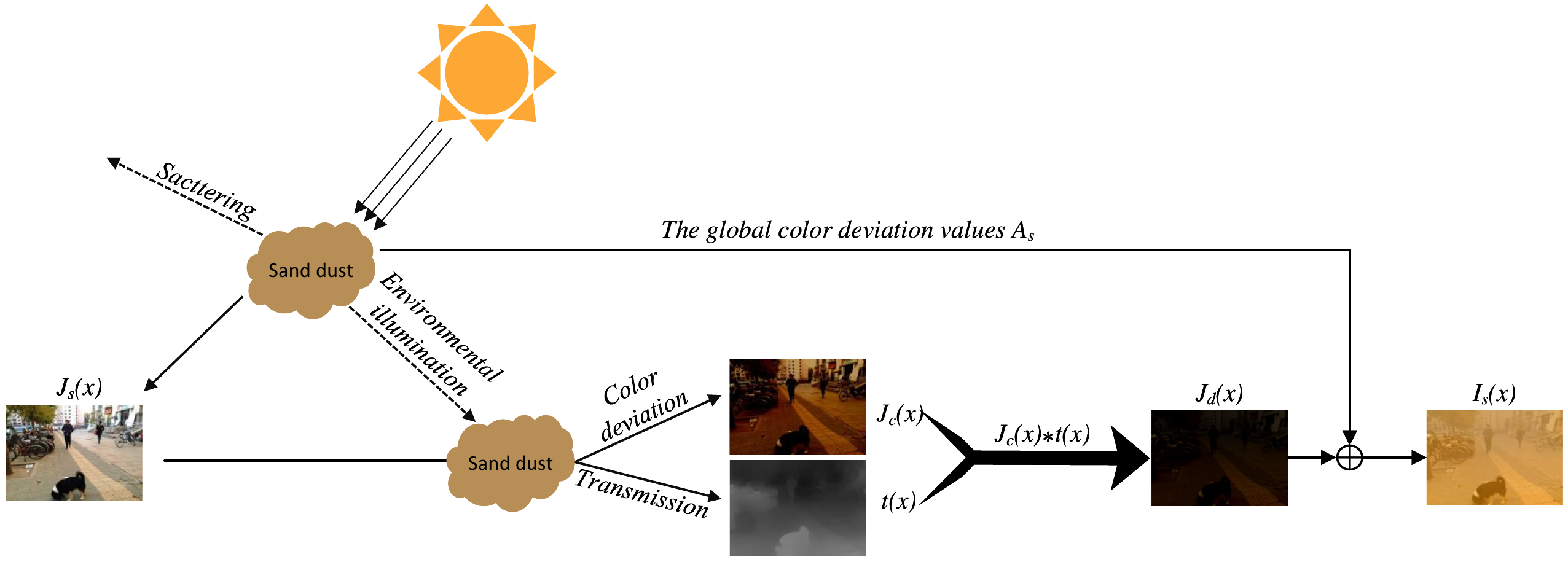}
	\caption{The flowchart of the proposed sand dust scattering model. Where $J_c(x)$ and $J_d(x)$ are the intermediate variables in the model. $J_c(x)$ represents the inherent color deviation of the object surface when the color deviation value is $A_s$. $J_d(x)$ is used to characterize the distribution of the dust in synthetic image.}
	\label{Fig.3}	
\end{figure*}

Actually, the performance of an algorithm cannot be fully evaluated only using the non-reference metrics. It is still necessary to calculate the full-reference metrics with paired samples. To our best knowledge, however, there is no publicly available dataset suitable for evaluating full-reference metrics of the sand dust removal algorithms up till now. Furthermore, the absence of the large-scale sandstorm dataset seriously hinders the development of data-driven based sand dust image reconstruction algorithms. 

\begin{table}
  \centering
  \caption{Statistics of the proposed SIRB}
  \setlength{\tabcolsep}{0.5mm}{
    \begin{tabular}{ccccc}
    \toprule
    \multicolumn{5}{c}{\textit{\textbf{Training Set}}} \\
    \textit{Subsets} & \textit{Number} & \textit{Dust intensity} & \textit{Category} & \textit{Attenuation coefficient} \\
    \midrule
    SIRB-T(L) & 3900 & Light & Synthetic & $\beta\in$[0.3,0.4] \\
    SIRB-T(M) & 3900 & Medium & Synthetic & $\beta\in$[0.4,0.5] \\
    SIRB-T(D) & 3900 & Dense & Synthetic & $\beta\in$[0.5,0.6] \\
    SIRB-T(H) & 3900 & Hybrid & Synthetic & $\beta\in$[0.3,0.6] \\
    \midrule
    \midrule
    \multicolumn{5}{c}{\textit{\textbf{Evaluating Set}}} \\
    \textit{Subsets} & \textit{Number} & \textit{Dust intensity} & \textit{Category} & \textit{Attenuation coefficient} \\
    \midrule
    SIRB-E(L) & 100  & Light & Synthetic & $\beta\in$[0.3,0.4] \\
    SIRB-E(M) & 100  & Medium & Synthetic & $\beta\in$[0.4,0.5] \\
    SIRB-E(D) & 100  & Dense & Synthetic & $\beta\in$[0.5,0.6] \\
    SIRB-E(H) & 100  & Hybrid & Synthetic & $\beta\in$[0.3,0.6] \\
    RSTS  & 230  & Hybrid & Real  & $\beta\in$[0,1] \\
    \bottomrule
    \end{tabular}}%
  \label{tab:1}%
\end{table}%

\section{Sand dust Image Reconstruction Benchmark}
\subsection{The characteristics of sand dust images}
The distribution of sand dust images has obvious prior characteristics for shifting, concentration, and sequential. Where shifting refers to that affected by Mai scattering, the R, G, B channels of sand dust images are dispersed; Concentration demonstrates that the pixel values in R, G, B channels are concentrated in a certain interval; Sequential refers to that the shifting distribution of R, G, B channels are in order according to G, B, R. We provide some histogram comparison examples of real-world sandstorm images and synthetic images, as shown in Figure \ref{Fig.2}, the distribution of synthetic images meets all properties of real-world sandstorm images.

\begin{figure}
	\centering
	\includegraphics[width=1\columnwidth]{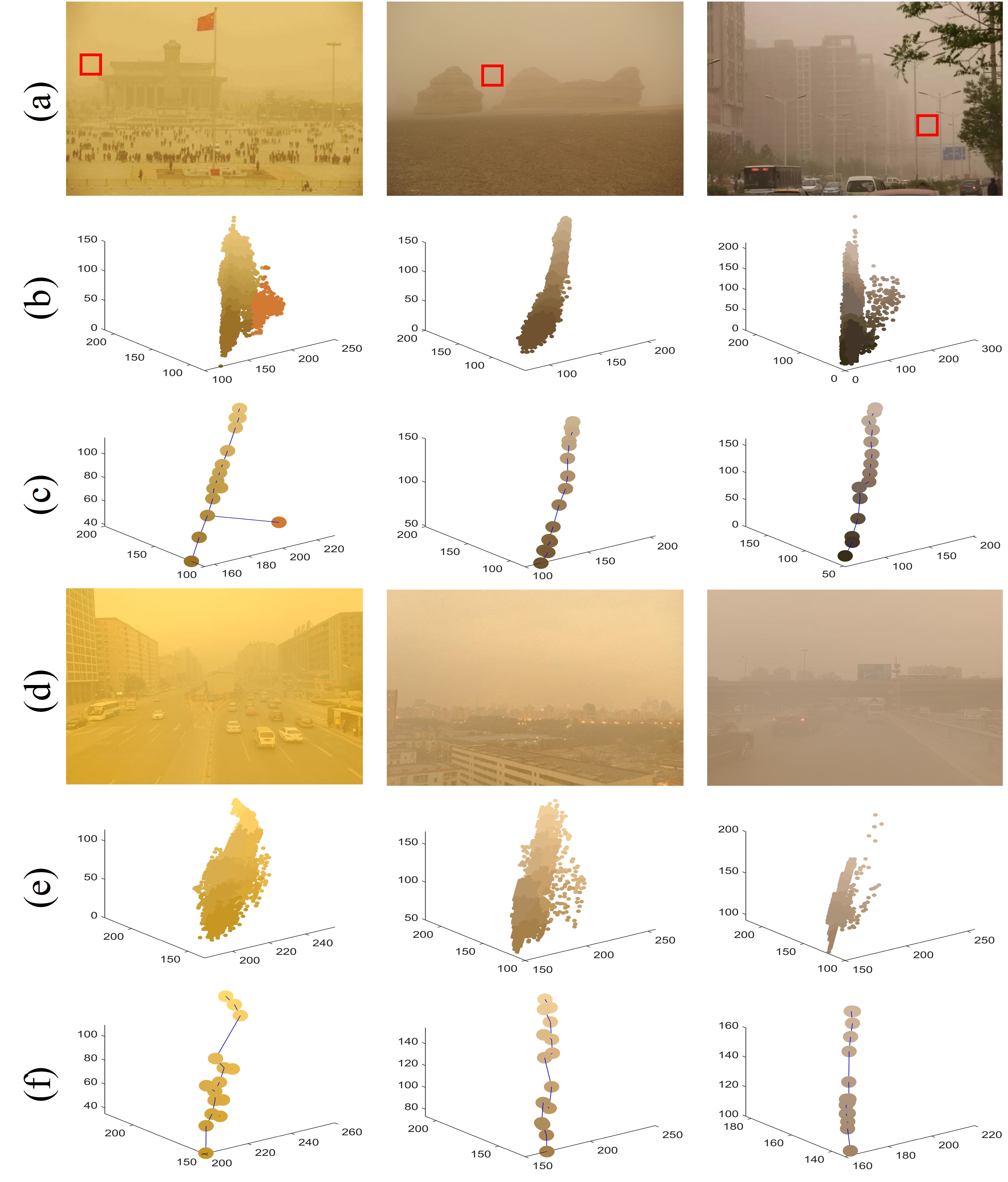}
	\caption{The color quantization and cluster of the sand dust images. (a) Real-world sand dust images; (b) Color quantization of real-world images; (c) Color cluster of real-world images; (d) Synthetic sand dust images; (e) Color quantization of synthetic images; (f) Color cluster of synthetic images. }
	\label{Fig.4}	
\end{figure}

\begin{figure*}
	\centering
	\includegraphics[width=2.1\columnwidth]{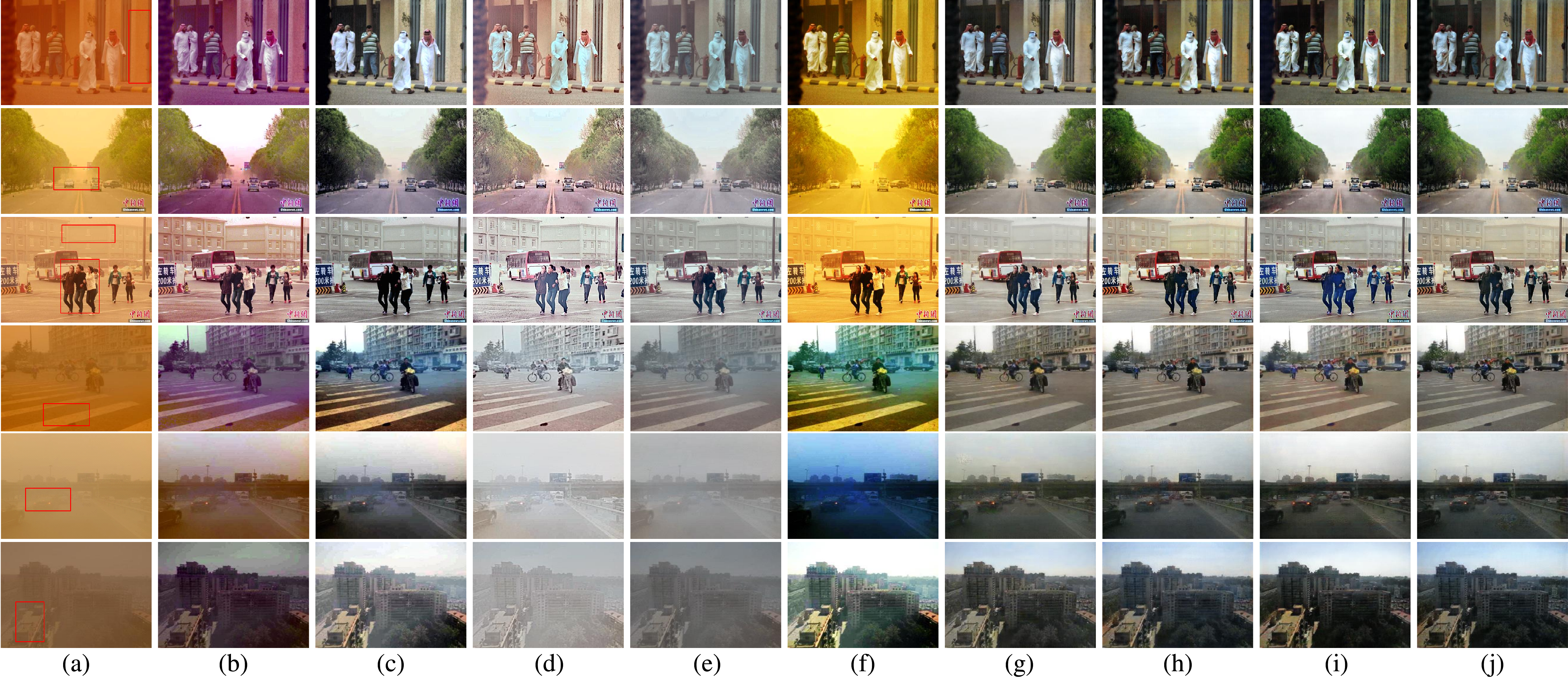}
	\caption{The comparisons on sand dust degraded images from SIRB. (a) Sand dust images; (b) CIDC \cite{ref2}; (c) FBE \cite{ref14}; (d) HRDCP \cite{ref4}; (e) NGT \cite{ref10}; (f) TTFIO \cite{ref58}; (g) Pix2pix-L \cite{ref19}; (h) Pix2pix-M \cite{ref19}; (i) Pix2pix-D \cite{ref19}; (j) Pix2pix-H \cite{ref19}.}
	\label{Fig.5}	
\end{figure*}

\subsection{The proposed synthesis strategies}
It is worth noting that under the sandstorm weather, the atmospheric scattering model does not hold any longer due to the radius of particles in sand dust is nearly 25$\mu$m \cite{ref55}, which is much larger than haze (0.01$\mu$m$\sim$1$\mu$m) and fog (1$\mu$m$\sim$10$\mu$m) \cite{ref56}. Considering the attenuation effect of sand dust floating in the atmosphere on RGB channels, we proposed a novel sand dust scattering model. The mathematical expression of the model is written as the following:

\begin{equation}
    {I_s}(x) = \left[ {{J_s}(x) - {{A'_s}}} \right]t(x) + {A_s}
    \label{eq:4}
\end{equation}
\begin{equation}
    t(x) = {e^{ - \beta d(x)}}  
    \label{eq:5}
\end{equation}
where $I_s(x)$ is the synthetic image; $J_s(x)$ is the corresponding reference image of $I_s(x)$; $A_s$ is the global color deviation value of the sand dust image; $A^{'}_s$ is the complementary color of $A_s$; $t(x)$ is the transmission of $J_s(x)$; $\beta$ is the attenuation coefficient of the dust, it represents the density of dust in the image; $d(x)$ is the scene depth. 

The flowchart of the proposed scattering model as shown in Figure \ref{Fig.3}. We adopted $J_c(x)$ to describe the influence of the reflected light scattered by sand dust particles to the pixels at the $x$ coordinate position on the image. $J_d(x)$ represents the distribution characteristics of dust on the basis of $J_c(x)$. The mathematical expression of them are written as:

\begin{equation}
    {J_c}\left( x \right) = \left[ {{J_s}\left( x \right) + {A_s} - 1} \right]  
    \label{eq:6}
\end{equation}
\begin{equation}
    {J_d}\left( x \right) = {J_c}\left( x \right) * t\left( x \right)
    \label{eq:7}
\end{equation}

The value of $A_s$ changes quite little in the same sand dust scene, so we assume it as a certain value. By applying $A_s$ at $x$, we can obtain the pixel's imaging in sandstorm weather.

The clear images in SIRB all selected from RESIDE \cite{ref36}. We carefully excluded those scenes with haze to ensure the original images are suitable enough for synthesizing sand dust images. In addition, due to the yellow sunlight may confuse the neural network, the scenes containing sunrise and sunset were also eliminated. Under the condition of ensuring the diversity of the scenes as much as possible, we totally screened 4000 clear images including traffic, buildings, streets, woods, etc. as original samples for the benchmark. 

We adopted \cite{ref57} to estimate the scene depth and altered the dust intensity by randomly selecting $\beta$ from the uniform distribution interval [0.3, 0.6]. Finally, we synthesized the training set SIRB-T and evaluating set SIRB-E, resulting in a total of 16000 paired synthetic sand dust images. To further evaluate the application value of the sand dust algorithms, we used the key words "sandstorm", "sand dust weather" and "sand dust" to retrieve images from the Internet ($bing.com$). And constructed a \emph{Real-world Sandstorm Testing Set (RSTS)}, which is composed of 230 real-world sand dust images. The structure of SIRB is shown in Table \ref{tab:1}, SIRB-T including four training subsets, which are named as SIRB-T(L), SIRB-T(M), SIRB-T(D), and SIRB-T(H), respectively; SIRB-E consists of four evaluation subsets and named as SIRB-E(L), SIRB-E(M), SIRB-E(D), and SIRB-E(H), respectively. Where “T” short for training and “E” denotes evaluation; “L”, “M”, “D”, and “H” represent the images with light, medium, dense and hybrid sand dust density, respectively.

\begin{table*}
  \centering
  \caption{Non-reference metrics on evaluation datasets}
    \begin{tabular}{ccccccccc}
    \toprule
    Metrics & CIDC \cite{ref2}  & FBE \cite{ref14}   & HRDCP \cite{ref4} & NGT \cite{ref10}   & TTFIO \cite{ref58} & Pix2pix \cite{ref19} & Reference images & Real-world sand dust images \\
    \midrule
    \multicolumn{9}{c}{SIRB-E(L)} \\
    \midrule
    NIQE↓ & \textcolor[rgb]{ 0,  .439,  .753}{4.3632} & \textcolor[rgb]{ 0,  .69,  .314}{4.3282} & 4.7565 & \textcolor[rgb]{ 1,  0,  0}{4.243} & 4.8254 & 4.9222 & \textbf{4.9142} & \multirow{3}[2]{*}{\textbf{\textbackslash{}}} \\
    SSEQ↓ & 13.897 & \textcolor[rgb]{ 0,  .439,  .753}{12.66} & \textcolor[rgb]{ 0,  .69,  .314}{10.8126} & 14.6643 & 14.706 & \textcolor[rgb]{ 1,  0,  0}{8.6928} & \textbf{14.2559} &  \\
    BIQI↓ & 28.8989 & \textcolor[rgb]{ 0,  .439,  .753}{27.1753} & 29.1914 & 30.4207 & \textcolor[rgb]{ 0,  .69,  .314}{26.2888} & \textcolor[rgb]{ 1,  0,  0}{20.0077} & \textbf{22.7995} &  \\
    \midrule
    \multicolumn{9}{c}{SIRB-E(M)} \\
    \midrule
    NIQE↓ & \textcolor[rgb]{ 0,  .439,  .753}{4.5131} & \textcolor[rgb]{ 1,  0,  0}{4.4494} & 4.7682 & \textcolor[rgb]{ 0,  .69,  .314}{4.4573} & 4.879 & 4.8446 & \textbf{4.9142} & \multirow{3}[2]{*}{\textbf{\textbackslash{}}} \\
    SSEQ↓ & 15.0067 & \textcolor[rgb]{ 0,  .439,  .753}{13.7058} & \textcolor[rgb]{ 0,  .69,  .314}{12.2437} & 16.1395 & 15.4367 & \textcolor[rgb]{ 1,  0,  0}{8.7683} & \textbf{14.2559} &  \\
    BIQI↓ & 30.8213 & 28.756 & \textcolor[rgb]{ 0,  .69,  .314}{25.7518} & 31.9037 & \textcolor[rgb]{ 0,  .439,  .753}{28.1029} & \textcolor[rgb]{ 1,  0,  0}{20.8456} & \textbf{22.7995} &  \\
    \midrule
    \multicolumn{9}{c}{SIRB-E(D)} \\
    \midrule
    NIQE↓ & \textcolor[rgb]{ 0,  .69,  .314}{4.8926} & \textcolor[rgb]{ 0,  .439,  .753}{4.9114} & 4.9507 & 4.923 & 5.3241 & \textcolor[rgb]{ 1,  0,  0}{4.8101} & \textbf{4.9142} & \multirow{3}[2]{*}{\textbf{\textbackslash{}}} \\
    SSEQ↓ & 15.8064 & \textcolor[rgb]{ 0,  .439,  .753}{14.7349} & \textcolor[rgb]{ 0,  .69,  .314}{14.6102} & 17.7692 & 16.4857 & \textcolor[rgb]{ 1,  0,  0}{8.8167} & \textbf{14.2559} &  \\
    BIQI↓ & 32.4881 & 30.6915 & \textcolor[rgb]{ 0,  .69,  .314}{24.6167} & 32.8375 & \textcolor[rgb]{ 0,  .439,  .753}{30.278} & \textcolor[rgb]{ 1,  0,  0}{21.6453} & \textbf{22.7995} &  \\
    \midrule
    \multicolumn{9}{c}{SIRB-E(H)} \\
    \midrule
    NIQE↓ & \textcolor[rgb]{ 0,  .439,  .753}{4.6019} & \textcolor[rgb]{ 1,  0,  0}{4.5309} & 4.8667 & \textcolor[rgb]{ 0,  .69,  .314}{4.5477} & 5.0053 & 4.8897 & \textbf{4.9142} & \multirow{3}[2]{*}{\textbf{\textbackslash{}}} \\
    SSEQ↓ & 15.0674 & \textcolor[rgb]{ 0,  .439,  .753}{13.7526} & \textcolor[rgb]{ 0,  .69,  .314}{12.8604} & 16.3487 & 15.6197 & \textcolor[rgb]{ 1,  0,  0}{8.6616} & \textbf{14.2559} &  \\
    BIQI↓ & 30.716 & 28.8654 & \textcolor[rgb]{ 0,  .69,  .314}{26.8795} & 31.4879 & \textcolor[rgb]{ 0,  .439,  .753}{28.1405} & \textcolor[rgb]{ 1,  0,  0}{20.8184} & \textbf{22.7995} &  \\
    \midrule
    \multicolumn{9}{c}{RSTS} \\
    \midrule
    NIQE↓ & 4.4634 & \textcolor[rgb]{ 0,  .439,  .753}{4.3783} & \textcolor[rgb]{ 0,  .69,  .314}{4.3576} & 4.5757 & 4.8264 & \textcolor[rgb]{ 1,  0,  0}{4.2971} & \multirow{3}[2]{*}{\textbf{\textbackslash{}}} & \textbf{8.5704} \\
    SSEQ↓ & 23.0548 & \textcolor[rgb]{ 0,  .439,  .753}{21.7397} & \textcolor[rgb]{ 0,  .69,  .314}{21.3578} & 24.269 & 27.312 & \textcolor[rgb]{ 1,  0,  0}{20.6379} &       & \textbf{27.4906} \\
    BIQI↓ & 33.8807 & \textcolor[rgb]{ 0,  .439,  .753}{30.7382} & \textcolor[rgb]{ 0,  .69,  .314}{26.834} & 32.9201 & 32.8246 & \textcolor[rgb]{ 1,  0,  0}{25.6253} &       & \textbf{28.3195} \\
    \bottomrule
    \end{tabular}%
  \label{tab:2}%
\end{table*}%

Figure \ref{Fig.4} shows the color quantization and cluster of the sand dust images, $A_s$ of the synthetic images were selected from the red rectangles of the real-world degraded images in the same column. As shown in Figure \ref{Fig.4}(b) and (e), we extracted the images hue through color quantization method and adopted Eq.\ref{eq:8} as the loss function of K-means to cluster the color distribution in LAB space. 
\begin{equation}
    L\left( {c,\mu } \right) = min\sum\limits_{i = 1}^N {{{\left\| {{x_i} - {\mu _{{c_i}}}} \right\|}^2}} 
    \label{eq:8}
\end{equation}
\begin{equation}
    argmi{n_k}{\left\| {{x_{i}} - \mu _k^t} \right\|^2} \to c_{i}^t
    \label{eq:9}
\end{equation}

\begin{equation}
    argmi{n_\mu }\sum\limits_{i:c_i^t = k}^N {{{\left\| {{x_i} - \mu } \right\|}^2}}  \to \mu _k^{t + 1}
    \label{eq:10}
\end{equation}
where $N$ is the pixel number; $x_i$ is the sample in LAB space; $\mu _{{c_i}}$ represents the center point corresponding to the cluster and we set the number of cluster to 15; $t$ is the number of iterations; $\mu$ is the average value of the current cluster. 

We repeatedly calculate Eq.\ref{eq:9} and Eq.\ref{eq:10}, and update the corresponding parameters until $L$ converges. The clustering results are shown in Figure \ref{Fig.4}(c) and (f). For sand dust images, the sand dust concentration and the color shift tend to increase with the depth of field, and the spatial distribution of color categories should be approximately a straight line. In Figure \ref{Fig.4}, synthetic results are similar to the real-world degraded images both in terms of statistical characteristics and visual perception.

To the best of our current knowledge, SIRB is the first large scale sand dust images benchmark dataset, which has great potential for training the CNNs and evaluating the performance of sand dust removal algorithms. 

\section{Evaluation and Discussion}
Due to the absence of reference images, most studies using few selected real-world images from internet to evaluate the performance of sand dust removal algorithms. However, that cannot provide a comprehensive assessment of the algorithm's performance. Based on the constructed SIRB, we evaluated 5 representative algorithms of sand dust removal including CIDC \cite{ref2}, FBE \cite{ref14}, HRDCP \cite{ref4}, NGT \cite{ref10}, TTFIO \cite{ref58}. For the fairness of comparison, all the codes of the algorithms are from the website of their authors. Furthermore, to illustrate the general applicability of SIRB for training CNNs, we trained Pix2Pix \cite{ref19} on SIRB-T(L), SIRB-T(M), SIRB-T(D) and SIRB-T(H) respectively as baseline, and marked them as Pix2pix-L, Pix2pix-M, Pix2pix-D and Pix2pix-H. In the following, we will compare the performance of the algorithms on SIRB-E and RSTS both qualitatively and quantitatively.

\begin{table*}
  \centering
  \caption{Full-reference metrics on evaluation datasets}
  \setlength{\tabcolsep}{3mm}{
    \begin{tabular}{cccccccc}
    \toprule
    Metrics & CIDC \cite{ref2}  & FBE \cite{ref14}   & HRDCP \cite{ref4} & NGT \cite{ref10}   & TTFIO \cite{ref58} & Pix2pix \cite{ref19} & Reference images \\
    \midrule
    \multicolumn{8}{c}{SIRB-E(L)} \\
    \midrule
    PSNR(dB)↑ & 15.9212 & \textcolor[rgb]{ 0,  .69,  .314}{20.1792} & 11.6168 & 14.8567 & \textcolor[rgb]{ 0,  .439,  .753}{18.4317} & \textcolor[rgb]{ 1,  0,  0}{24.5471} & $+\infty$ \\
    MSE($\times10^2$)↓ & 19.4965 & \textcolor[rgb]{ 0,  .69,  .314}{7.3057} & 47.7826 & 23.803 & \textcolor[rgb]{ 0,  .439,  .753}{11.1207} & \textcolor[rgb]{ 1,  0,  0}{2.6383} & 0 \\
    SSIM↑ & 0.5369 & \textcolor[rgb]{ 0,  .69,  .314}{0.7292} & 0.575 & \textcolor[rgb]{ 0,  .439,  .753}{0.6614} & 0.5754 & \textcolor[rgb]{ 1,  0,  0}{0.8177} & 1 \\
    FSIMc↑ & 0.9088 & \textcolor[rgb]{ 0,  .69,  .314}{0.9225} & 0.8109 & 0.8994 & \textcolor[rgb]{ 0,  .439,  .753}{0.916} & \textcolor[rgb]{ 1,  0,  0}{0.9322} & 1 \\
    CIE94↓ & 21.0675 & \textcolor[rgb]{ 0,  .69,  .314}{10.8067} & 25.0726 & \textcolor[rgb]{ 0,  .439,  .753}{17.5628} & 17.8953 & \textcolor[rgb]{ 1,  0,  0}{7.0081} & 0 \\
    CIEDE2000↓ & 39.2437 & \textcolor[rgb]{ 0,  .69,  .314}{23.5664} & 29.9438 & \textcolor[rgb]{ 0,  .439,  .753}{25.9552} & 40.7036 & \textcolor[rgb]{ 1,  0,  0}{18.1497} & 0 \\
    \midrule
    \multicolumn{8}{c}{SIRB-E(M)} \\
    \midrule
    PSNR(dB)↑ & 15.3509 & \textcolor[rgb]{ 0,  .69,  .314}{18.9701} & 10.7092 & 13.8026 & \textcolor[rgb]{ 0,  .439,  .753}{17.0033} & \textcolor[rgb]{ 1,  0,  0}{23.9323} & $+\infty$ \\
    MSE($\times10^2$)↓ & 21.996 & \textcolor[rgb]{ 0,  .69,  .314}{9.822} & 59.0351 & 29.9207 & \textcolor[rgb]{ 0,  .439,  .753}{16.2594} & \textcolor[rgb]{ 1,  0,  0}{3.0561} & 0 \\
    SSIM↑ & 0.5073 & \textcolor[rgb]{ 0,  .69,  .314}{0.6811} & 0.5536 & \textcolor[rgb]{ 0,  .439,  .753}{0.5962} & 0.4978 & \textcolor[rgb]{ 1,  0,  0}{0.7935} & 1 \\
    FSIMc↑ & 0.8853 & \textcolor[rgb]{ 0,  .69,  .314}{0.9068} & 0.8189 & 0.8431 & \textcolor[rgb]{ 0,  .439,  .753}{0.8924} & \textcolor[rgb]{ 1,  0,  0}{0.9208} & 1 \\
    CIE94↓ & 21.8312 & \textcolor[rgb]{ 0,  .69,  .314}{12.2343} & 27.7036 & \textcolor[rgb]{ 0,  .439,  .753}{19.5966} & 19.6172 & \textcolor[rgb]{ 1,  0,  0}{7.2978} & 0 \\
    CIEDE2000↓ & 40.1243 & \textcolor[rgb]{ 0,  .69,  .314}{24.3841} & 30.6187 & \textcolor[rgb]{ 0,  .439,  .753}{27.0389} & 39.839 & \textcolor[rgb]{ 1,  0,  0}{18.2889} & 0 \\
    \midrule
    \multicolumn{8}{c}{SIRB-E(D)} \\
    \midrule
    PSNR(dB)↑ & 14.8224 & \textcolor[rgb]{ 0,  .69,  .314}{17.8168} & 9.9632 & 13.0396 & \textcolor[rgb]{ 0,  .439,  .753}{15.5809} & \textcolor[rgb]{ 1,  0,  0}{23.0025} & $+\infty$ \\
    MSE($\times10^2$)↓ & 24.3192 & \textcolor[rgb]{ 0,  .69,  .314}{12.6623} & 69.9469 & 35.1292 & \textcolor[rgb]{ 0,  .439,  .753}{23.1261} & \textcolor[rgb]{ 1,  0,  0}{3.8136} & 0 \\
    SSIM↑ & 0.476 & \textcolor[rgb]{ 0,  .69,  .314}{0.6291} & 0.5194 & \textcolor[rgb]{ 0,  .439,  .753}{0.5381} & 0.434 & \textcolor[rgb]{ 1,  0,  0}{0.7609} & 1 \\
    FSIMc↑ & 0.8593 & \textcolor[rgb]{ 0,  .69,  .314}{0.8855} & 0.8065 & 0.7888 & \textcolor[rgb]{ 0,  .439,  .753}{0.8659} & \textcolor[rgb]{ 1,  0,  0}{0.9064} & 1 \\
    CIE94↓ & 22.5045 & \textcolor[rgb]{ 0,  .69,  .314}{13.8246} & 29.9714 & \textcolor[rgb]{ 0,  .439,  .753}{21.2139} & 21.5307 & \textcolor[rgb]{ 1,  0,  0}{7.9267} & 0 \\
    CIEDE2000↓ & 40.6899 & \textcolor[rgb]{ 0,  .69,  .314}{25.9296} & 31.2957 & \textcolor[rgb]{ 0,  .439,  .753}{27.9186} & 38.8379 & \textcolor[rgb]{ 1,  0,  0}{18.988} & 0 \\
    \midrule
    \multicolumn{8}{c}{SIRB-E(H)} \\
    \midrule
    PSNR(dB)↑ & 15.4827 & \textcolor[rgb]{ 0,  .69,  .314}{18.9324} & 10.8071 & 13.9506 & \textcolor[rgb]{ 0,  .439,  .753}{16.9514} & \textcolor[rgb]{ 1,  0,  0}{23.9422} & $+\infty$ \\
    MSE($\times10^2$)↓ & 21.3386 & \textcolor[rgb]{ 0,  .69,  .314}{10.111} & 58.1206 & 28.9367 & \textcolor[rgb]{ 0,  .439,  .753}{17.008} & \textcolor[rgb]{ 1,  0,  0}{3.039} & 0 \\
    SSIM↑ & 0.5097 & \textcolor[rgb]{ 0,  .69,  .314}{0.6799} & 0.5477 & \textcolor[rgb]{ 0,  .439,  .753}{0.6025} & 0.4989 & \textcolor[rgb]{ 1,  0,  0}{0.7926} & 1 \\
    FSIMc↑ & 0.8869 & \textcolor[rgb]{ 0,  .69,  .314}{0.906} & 0.8108 & 0.8476 & \textcolor[rgb]{ 0,  .439,  .753}{0.8937} & \textcolor[rgb]{ 1,  0,  0}{0.9204} & 1 \\
    CIE94↓ & 21.5635 & \textcolor[rgb]{ 0,  .69,  .314}{12.3904} & 27.4295 & \textcolor[rgb]{ 0,  .439,  .753}{19.2932} & 19.6119 & \textcolor[rgb]{ 1,  0,  0}{7.3264} & 0 \\
    CIEDE2000↓ & 39.8556 & \textcolor[rgb]{ 0,  .69,  .314}{24.7188} & 30.7003 & \textcolor[rgb]{ 0,  .439,  .753}{26.9774} & 39.6975 & \textcolor[rgb]{ 1,  0,  0}{18.2764} & 0 \\
    \bottomrule
    \end{tabular}}%
  \label{tab:3}%
\end{table*}%

\subsection{Qualitative evaluation}
As shown in Figure \ref{Fig.5}, We provided several synthetic and real-word sand dust degraded images from SIRB as the comparison samples, please zoom in for a better illustration. The color correction ability of CIDC \cite{ref2} is the weakest, followed by TTFIO \cite{ref58}, the results of them have obvious color distortion issues. In the comparisons, HRDCP \cite{ref4} and NGT \cite{ref10} can effectively correct the color, but the noise was amplified as well, there still existing residue dust in the images. For FBE \cite{ref14}, it can effectively correct the color and improve the clarity of the images, but the results are slightly dark and with low saturation. In Figure \ref{Fig.5}, The results of the CNNs are visually more pleasing, as shown in the red marked areas, the color correction ability of Pix2pix-L is stronger than the other three training strategies, but the dust removal ability is relatively weak, the area in the deep scenes of the results are still hazy; While the dust removal ability of Pix2pix-D is better than the other three training strategies, but it is weak in term of color correction, the local area still exist the issue of color shift. Moreover, in the local area with high dust concentration, Pix2pix may loss the detail information, it is a noteworthy issue in future sand dust image reconstruction research.

In the comparison experiment, we found that the performance of CNNs is affected by the training dataset to a large extent. Pix2pix-L can correct the color effectively, but the background is still hazy. On the contrary, Pix2pix-D has stronger dust removal ability, but there is slight color distortion in local areas. Overall, the Pix2pix-M and Pix-pix-H can keep better balance between color correction and dust removal. Scholars should choose different training strategies according to the demands of their sand dust removal tasks.

\subsection{Quantitative evaluation}
\subsubsection{Non-reference evaluation}
We tested the performance of the algorithms on synthetic evaluation datasets and RSTS with three non-reference metrics including NIQE, SSEQ, and BIQI. The average scores are shown in Table \ref{tab:2}, where the lower metrics indicate the less distortion and more natural of the results.

In Table \ref{tab:2}, the best performance, the second best and the third best are highlighted in red, green, and blue respectively. The metrics of Pix2pix are the average values of the CNNs trained on four subsets. Clearly, compared with the traditional algorithms, Pix2pix has the absolute advantage in the performance of indicators both synthetic datasets and RSTS. In addition, we evaluated the image quality of real-world sand dust degraded images and reference images, and marked them in bold. It is undeniable that though the non-reference evaluation algorithms had claimed they took the human visual perception into account, but the metrics of the reference images provide evidence that the evaluation algorithms without reference may not always be accurate. The scores of reference images are even higher than most results of the comparison algorithms. In addition, for the algorithms tested on RSTS, the scores of their results are even higher than original degraded images, which are unfair evaluations from the perspective of human visual perception. Although the non-reference metrics can rank the algorithms performance to some extent, they are intrinsically limited due to the unavailability to comprehensive study and analyze the results, especially their poor alignment with human perception for the images with extremely serious color deviation in such sandstorm environment.

\begin{table*}
  \centering
  \caption{Comparisons of time complexity (in second)}
  \setlength{\tabcolsep}{3mm}{
    \begin{tabular}{ccccccc}
    \toprule
    Image sizes & CIDC \cite{ref2}  & FBE \cite{ref14}   & HRDCP \cite{ref4} & NGT \cite{ref10}   & TTFIO \cite{ref58} & Pix2pix \cite{ref19} \\
    \midrule
    256×256 & 0.1356 & 0.0891 & 0.4268 & \textcolor[rgb]{ 0,  .439,  .753}{0.037} & \textcolor[rgb]{ 0,  .69,  .314}{0.016} & \textcolor[rgb]{ 1,  0,  0}{0.0148} \\
    512×512 & 0.6493 & 0.2957 & 1.6464 & \textcolor[rgb]{ 0,  .439,  .753}{0.111} & \textcolor[rgb]{ 0,  .69,  .314}{0.0837} & \textcolor[rgb]{ 1,  0,  0}{0.0235} \\
    1024×1024 & 2.8163 & 1.1904 & 6.767 & \textcolor[rgb]{ 0,  .69,  .314}{0.4032} & \textcolor[rgb]{ 0,  .439,  .753}{0.675} & \textcolor[rgb]{ 1,  0,  0}{0.0691} \\
    \bottomrule
    \end{tabular}}%
  \label{tab:4}%
\end{table*}%

\subsubsection{Full-reference evaluation}
To some extent, by using the reference images, the metrics of full-reference evaluation strategy can provide realistic feedback of the comparison algorithms, although there might be differences between the reference images and the real ground truth images. To complement the one-sided of the non-reference evaluation, we evaluated the algorithms by calculating pixel error, structural error, and color deviation.

Pixel error is one of the most intuitive criteria for people to perceive image quality, we employed two most popular indicators PSNR and MSE to measure the pixel error. As shown in Table \ref{tab:3}, we present the average metrics of the algorithms on the synthetic evaluation datasets, the result with higher PSNR and lower MSE scores means it is closer to the reference image in term of content. In Table \ref{tab:3}, Pix2pix \cite{ref19} has the best overall performance, and FBE \cite{ref14} ranks the second-best overall performance in pixel error, followed by TTFIO \cite{ref58}. As shown in Figures \ref{Fig.1} and \ref{Fig.5}, CIDC \cite{ref2} and HRDCP \cite{ref4} amplified the noise when removing sand and dust, leading to the larger pixel error in the results. 

We adopted SSIM and FSIMc to measure the structural information of the results, which are widely used to evaluate the structure differences between two images. The higher SSIM and FSIMc scores denote the structure of the results are closer to the reference images. We measured the structural errors of the algorithms on four evaluation subsets, and we also provide the metrics of the reference images as the target line. In Table \ref{tab:3}, as the traditional algorithms, FBE \cite{ref14} has the second best overall performance, followed by TTFIO \cite{ref58}, but there is still an obvious gap between the traditional algorithms and CNNs based method. 

The most intuitive feature of the sand dust images is color shift. Whether the issue of color distortion can be properly handled is directly reflect the color correction ability of the algorithms, which also an important reference to evaluate the performance of the algorithms. To refine the evaluation strategies, we supplemented CIE94 and CIEDE2000 to measure the color correction ability of the algorithms. The average metrics of the algorithms on the synthetic evaluation sets as shown in Table \ref{tab:3}, the lower CIE94 and CIEDE2000 metrics mean the result is closer to the reference image in term of color consistency. In Figures \ref{Fig.1} and \ref{Fig.5}, the results of CIDC \cite{ref2} and TTFIO \cite{ref58} still exist the issue of color distortion, that is the reason why both of them have higher CIE94 and CIEDE2000 metrics. Compared with traditional algorithms, the color correction ability of CNNs based method still shows the clear advantage. The experiments demonstrate that the learning based algorithms have great potential in sand dust removal task. 

\subsubsection{Execution time}
We evaluated the execution time of the algorithms by using Matlab2016a on a PC with an Intel(R) Core (TM) i5-9400 CPU@2.90GHz, 16GB RAM and network on a NVIDIA GeForce RTX2070 GPU. For fairness comparison, all the codes of the algorithms are from the website of their authors. As shown in Table \ref{tab:4}, we compared the average runtime for the sand dust images with different sizes. The running time of Pix2pix \cite{ref19} was taken from the average execution time of Pix2pix-L, Pix2pix-M, Pix2pix-D and Pix2pix-H. In Table \ref{tab:4}, NGT \cite{ref10}, TTFIO \cite{ref58} and Pix2pix can meet the requirements of real-time processing when handling 256$\times$256 images. With the increase of image size, the processing speed of the algorithms became slower, especially for traditional algorithms. The most time-consuming algorithm is HRDCP \cite{ref4}, the average running time of it is 6.767s when handling 1024$\times$1024 images. The processing speed of Pix2pix on 1024$\times$1024 images is about 15FPS, that means when the images size is less than 1024$\times$1024, Pix2pix can theoretically meet the needs of real-time processing in practical deployment.

\section{Conclusion and Future Research}
In this paper, we constructed a novel large-scale dataset named SIRB, which provide a supervised way to benchmark the performance of the algorithms. The dataset enabled us to conduct a thorough survey of the current sand dust removal algorithms. Moreover, we trained Pix2pix as baseline on SIRB-T and discussed the impact of training dataset for the performance of the network. Extensive experiments demonstrate that the data-driven based method superiority over the traditional algorithms in terms of qualitative and quantitative evaluation. To further compare the practical value of the algorithms, we implemented the time complexity analysis experiment. The experiment demonstrate that data-driven based method has an absolute advantage in execution time which shows the great potential of CNNs in the deployment of sand dust video processing. Several overarching observations and empirical findings are summarized as:
\begin{itemize}
\item Pix2pix trained on SIRB-T can remove the sandstorm to some extent, however, in the areas with dense sand dust concentration, the intensity of detail information is too weak, which may cause local information distorted in the reconstruction images. How to better preserve the detail information in sand dust removal task needs further research and discussion.
\item In CNNs, the appropriate prior knowledge may play a positive role in training. We advocate a combination of appropriate priors and data-driven algorithms to make their advantages complementary.
\item In the execution time experiment, we found that besides processing the single image, the data-driven method also has the potential for video processing. It is possible to handle high-resolution images in real time by cutting redundant branches and deleting unnecessary parameters of the network.
\end{itemize}

Although the proposed method can generate visually realistic sand dust images, based on the experiments and analysis there still some issues cannot be ignored. Specifically, due to the inaccurate estimation of depth, some synthetic images have dense sand dust concentration in the foreground and low concentration in the background; In addition, how to expand the sand dust intensity and environment types more suitable for training CNNs are still the focus in our future research.

\begin{acks}
This work is supported by National Key Research and Development Program of China(2019YFB1312102) and Natural Science Foundation of Hebei Province(F2019202364).
\end{acks}

\appendix

\section{A Comprehensive Benchmark Analysis for Sand Dust Image Reconstruction: Supplementary Material}

\subsection{Further study of the sand dust degraded images feature}
\subsubsection{Irreversible sand dust distribution}
\begin{figure}[htbp]
 \centering
 \includegraphics[width=1\columnwidth]{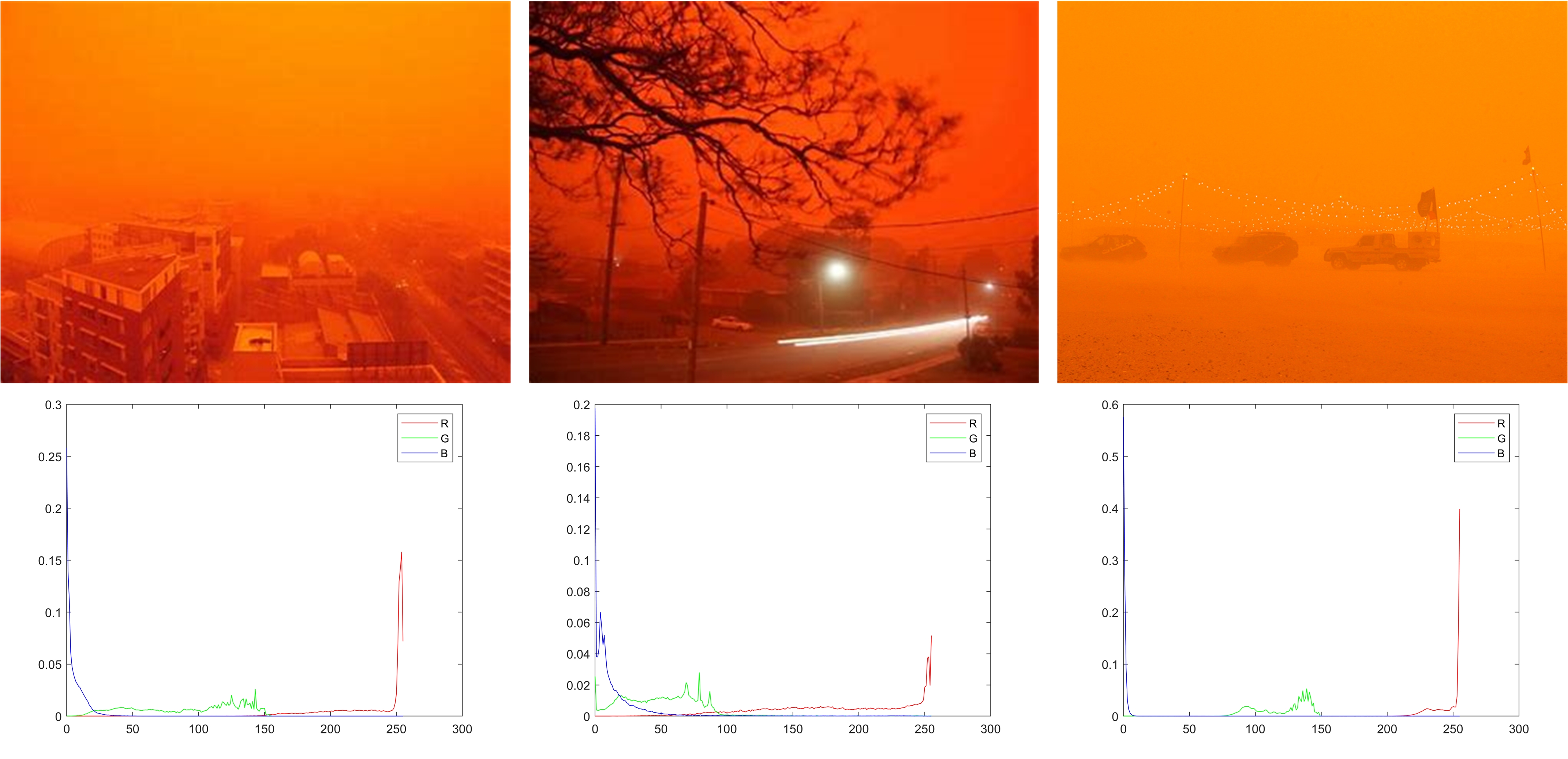}
 \caption{In the villainous sandstorm environment, extensive color information of the image is lost.}
 \label{Fig.6} 
\end{figure}

To improve the quality of the dataset, we have done much research on the color deviation distribution of real-world sand dust images. Almost all sand dust images retain complete RGB channel information except for the few images with extremely serious color deviation. As shown in Figure \ref{Fig.6}, a large amount of color information was lost due to the extremely serious color shifting. The energy of B is almost 0, while the energy of R is overflows to 255, that situations are irreversible, so we will not further analyze and discuss it in the article. 

\subsection{More results synthesized by the proposed method}
\subsubsection{Controllable synthetic results generated by proposed method}
In SIRB, the hyper-parameters $A_s$ and $\beta$, can be changed to generate controllable synthetic sand dust images. The sandstorm environment is too complex, the diversity of color deviation and dust intensity could play a positive role in training neural networks with better generalization ability.

\textbf{Global color deviation values} To improve the diversity and authenticity of the dataset, we carried out empirical research on real-world sandstorm images to obtain the sand distribution and all the global color deviation values of the synthetic images are taken from the area of maximum depth in real-world sandstorm images. The global color deviation values of SIRB are shown in Table \ref{tab:5}, and the values all following the principle of $R$\textgreater$G$\textgreater$B$. To avoid the loss of color information due to the underflow or overflow of pixels, the global color deviation values in SIRB all can limit the original information to a valid interval.

\begin{figure*}
 \centering
 \includegraphics[width=1.98\columnwidth]{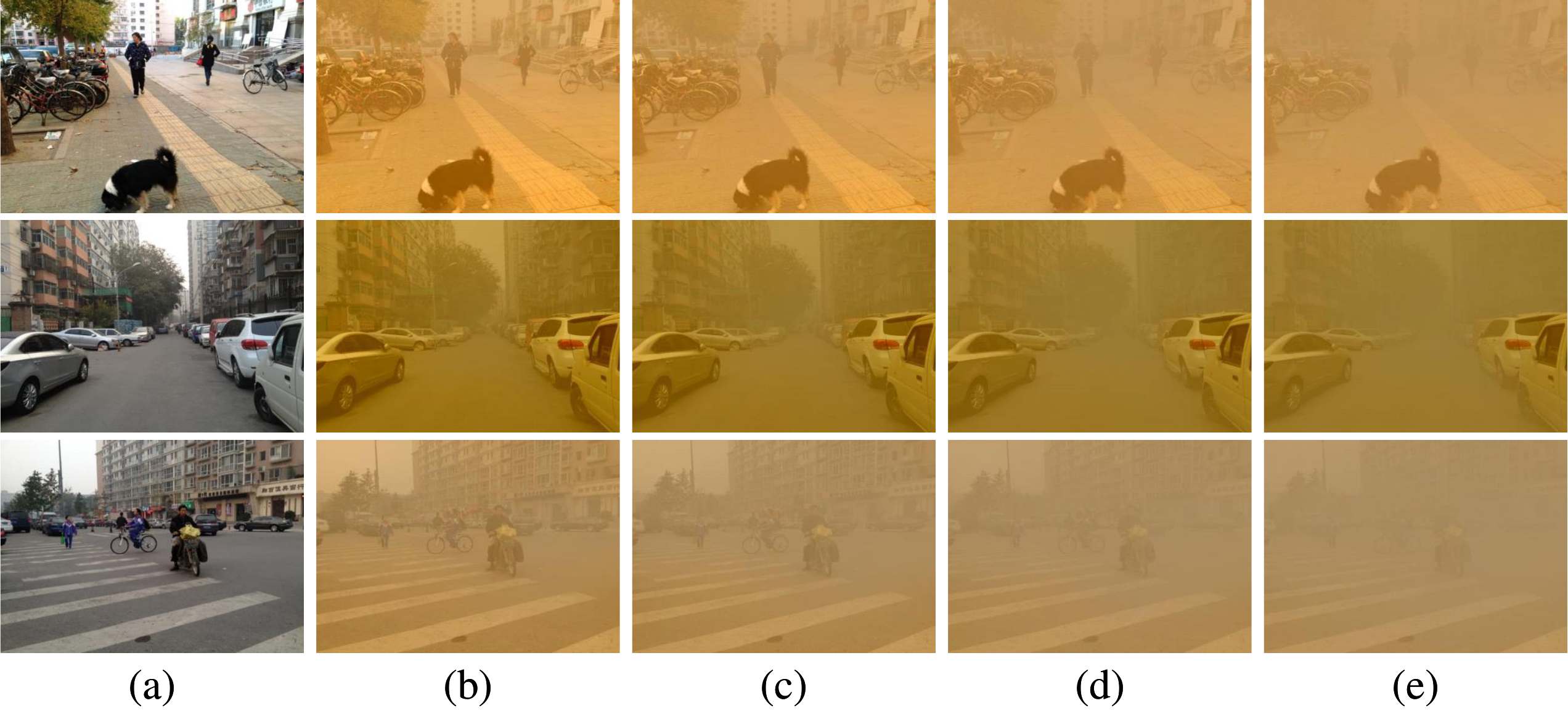}
 \caption{The synthetic results with different sand dust density levels. (a) Clear images; (b) $\beta$=0.3; (c) $\beta$=0.4; (d) $\beta$=0.5; (e) $\beta$=0.6.}
 \label{Fig.7} 
\end{figure*}

\textbf{Sand dust intensity}
To improve the quality of the synthetic images, we set the lower and upper limits of $\beta$ to 0.3 and 0.6, respectively. As shown in Figure \ref{Fig.7}, we provide three synthetic samples to illustrate the effect of $\beta$ on the synthesis results. With the increasing of $\beta$, the concentration of sand dust is higher, and the images become more hazy.

\subsubsection{Visual results on public datasets}
\begin{figure*}
 \centering
 \includegraphics[width=1.98\columnwidth]{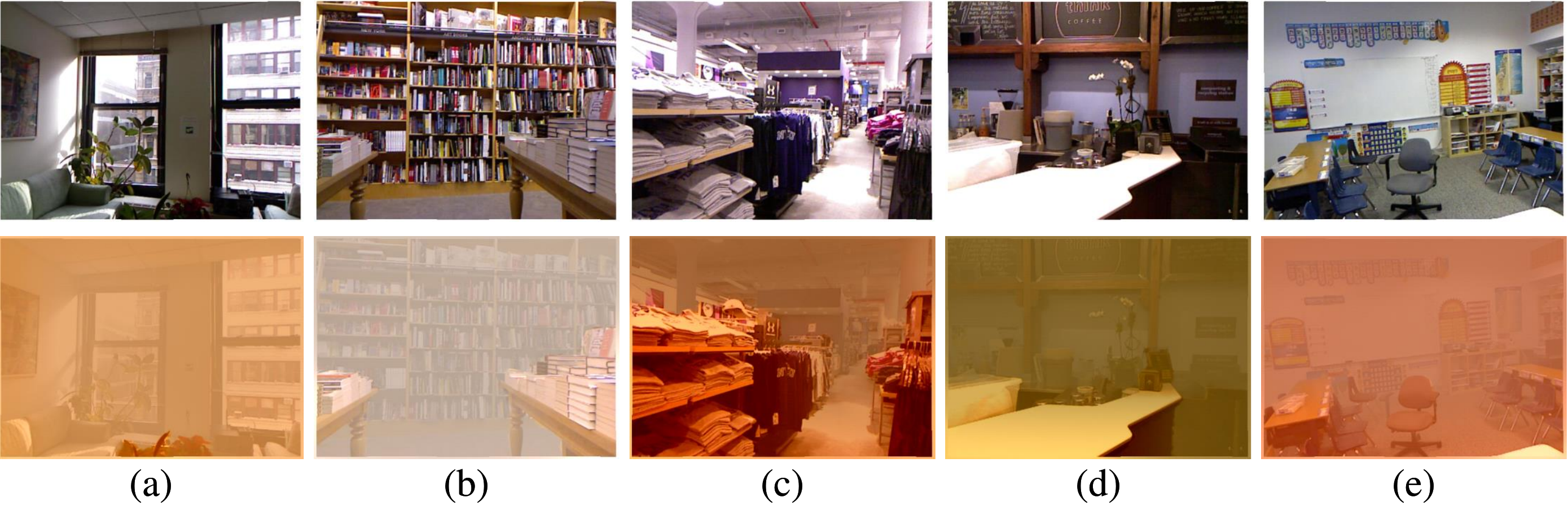}
 \caption{The synthetic results based on NYU \cite{ref59}, the top row are the reference images and the bottom row are the synthetic images. The corresponding color hex from left to right are "$\#C89463$", "$\#B9A99C$", "$\#986339$", "$\#826934$" and "$\#B97455$".}
 \label{Fig.8} 
\end{figure*}

\begin{figure*}
 \centering
 \includegraphics[width=1.98\columnwidth]{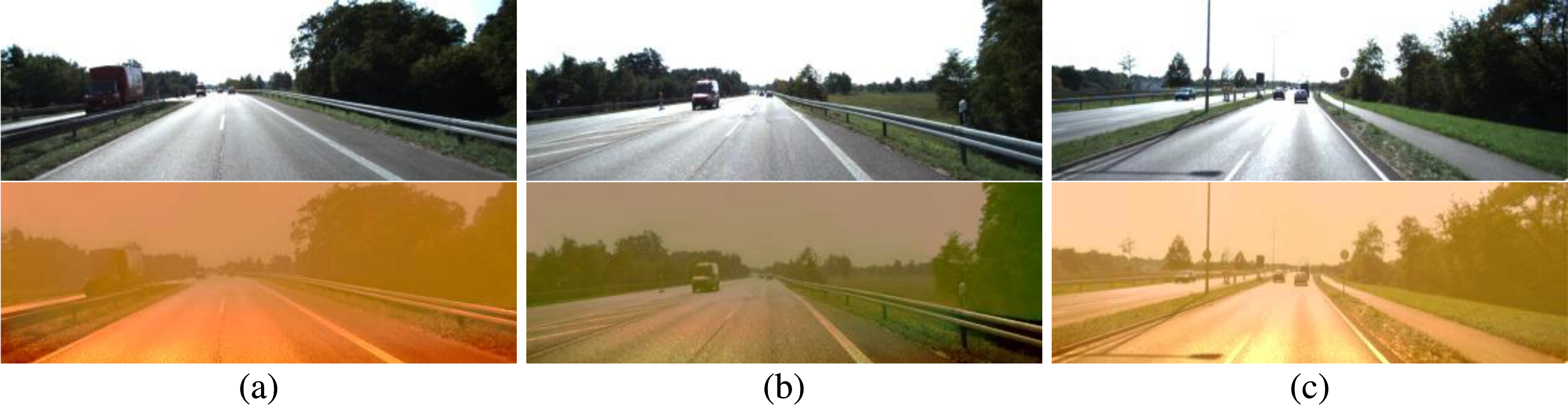}
 \caption{The synthetic results based on KITTI \cite{ref60}, the top row are the reference images and the bottom row are the synthetic images. The corresponding color hex from left to right are "$\#B37A43$", "$\#7D6E4A$" and "$\#B39163$".}
 \label{Fig.9} 
\end{figure*}

We synthesized the sand dust images using NYU \cite{ref59} and KITTI \cite{ref60} datasets to further illustrate the generality of the proposed synthesis method. As shown in Figures \ref{Fig.8} and \ref{Fig.9}, the proposed synthesis method can simulate the characteristics of the sandstorm degraded image both indoor and outdoor scene. That means we can synthesize more valuable sand dust images based on the public datasets for training neural network and evaluating the performance of the sand dust removal algorithms.
\begin{table}
  \centering
  \caption{The global color deviation values using in SIRB}
    \setlength{\tabcolsep}{2mm}{
    \begin{tabular}{cccccc}
    \toprule
         \makebox[0.001\textwidth][c]{} & \makebox[0.02\textwidth][c]{Color} & \makebox[0.08\textwidth][c]{R} & \makebox[0.08\textwidth][c]{G} & \makebox[0.08\textwidth][c]{B} & \makebox[0.02\textwidth][c]{Hex} \\
    \midrule
          & \cellcolor[rgb]{ .784,  .58,  .388} & 0.78431 & 0.58039 & 0.38824     & \#C89463 \\\addlinespace[0.5ex]
          & \cellcolor[rgb]{ .776,  .522,  .239} & 0.77647 & 0.52157 & 0.23922     & \#C6853D \\\addlinespace[0.5ex]
          & \cellcolor[rgb]{ .757,  .443,  .216} & 0.75686 & 0.44314 & 0.21569     & \#C17137 \\\addlinespace[0.5ex]
          & \cellcolor[rgb]{ .733,  .612,  .529} & 0.73333 & 0.61176 & 0.52941    & \#BB9C87 \\\addlinespace[0.5ex]
          & \cellcolor[rgb]{ .725,  .663,  .612} & 0.72549 & 0.66275 & 0.61176    & \#B9A99C \\\addlinespace[0.5ex]
          & \cellcolor[rgb]{ .725,  .455,  .333} & 0.72549 & 0.4549 & 0.33333     & \#B97455 \\\addlinespace[0.5ex]
          & \cellcolor[rgb]{ .718,  .557,  .337} & 0.71765 & 0.55686 & 0.33725     & \#B78E56 \\\addlinespace[0.5ex]
          & \cellcolor[rgb]{ .71,  .435,  .294} & 0.7098 & 0.43529 & 0.29412     & \#B56F4B \\\addlinespace[0.5ex]
          & \cellcolor[rgb]{ .702,  .478,  .263} & 0.70196 & 0.47843 & 0.26275     & \#B37A43 \\\addlinespace[0.5ex]
          & \cellcolor[rgb]{ .702,  .569,  .388} & 0.70196 & 0.56863 & 0.38824     & \#B39163 \\\addlinespace[0.5ex]
          & \cellcolor[rgb]{ .655,  .443,  .208} & 0.6549 & 0.44314 & 0.20784     & \#A77135 \\\addlinespace[0.5ex]
          & \cellcolor[rgb]{ .647,  .537,  .38} & 0.64706 & 0.53725 & 0.38039     & \#A58961 \\\addlinespace[0.5ex]
          & \cellcolor[rgb]{ .631,  .29,  .063} & 0.63137 & 0.2902 & 0.06275     & \#A14A10 \\\addlinespace[0.5ex]
          & \cellcolor[rgb]{ .596,  .388,  .224} & 0.59608 & 0.38824 & 0.22353     & \#986339 \\\addlinespace[0.5ex]
          & \cellcolor[rgb]{ .592,  .478,  .22} & 0.59216 & 0.47843 & 0.21961     & \#977A38 \\\addlinespace[0.5ex]
          & \cellcolor[rgb]{ .549,  .431,  .22} & 0.54902 & 0.43137 & 0.21961     & \#8C6E38 \\\addlinespace[0.5ex]
          & \cellcolor[rgb]{ .518,  .404,  .298} & 0.51765 & 0.40392 & 0.29804     & \#84674C \\\addlinespace[0.5ex]
          & \cellcolor[rgb]{ .51,  .412,  .204} & 0.5098 & 0.41176 & 0.20392     & \#826934 \\\addlinespace[0.5ex]
          & \cellcolor[rgb]{ .49,  .431,  .29} & 0.4902 & 0.43137 & 0.2902     & \#7D6E4A \\\addlinespace[0.5ex]
          & \cellcolor[rgb]{ .486,  .463,  .416} & 0.48627 & 0.46275 & 0.41569    & \#7C766A \\\addlinespace[0.5ex]
          & \cellcolor[rgb]{ .435,  .337,  .2} & 0.43529 & 0.33725 & 0.2     & \#6F5633 \\
    \bottomrule
    \end{tabular}}%
  \label{tab:5}%
\end{table}%
\subsection{More comparisons of sand dust removal algorithms}
\subsubsection{Visual comparisons on SIRB}
\begin{figure*}
 \centering
 \includegraphics[width=2\columnwidth]{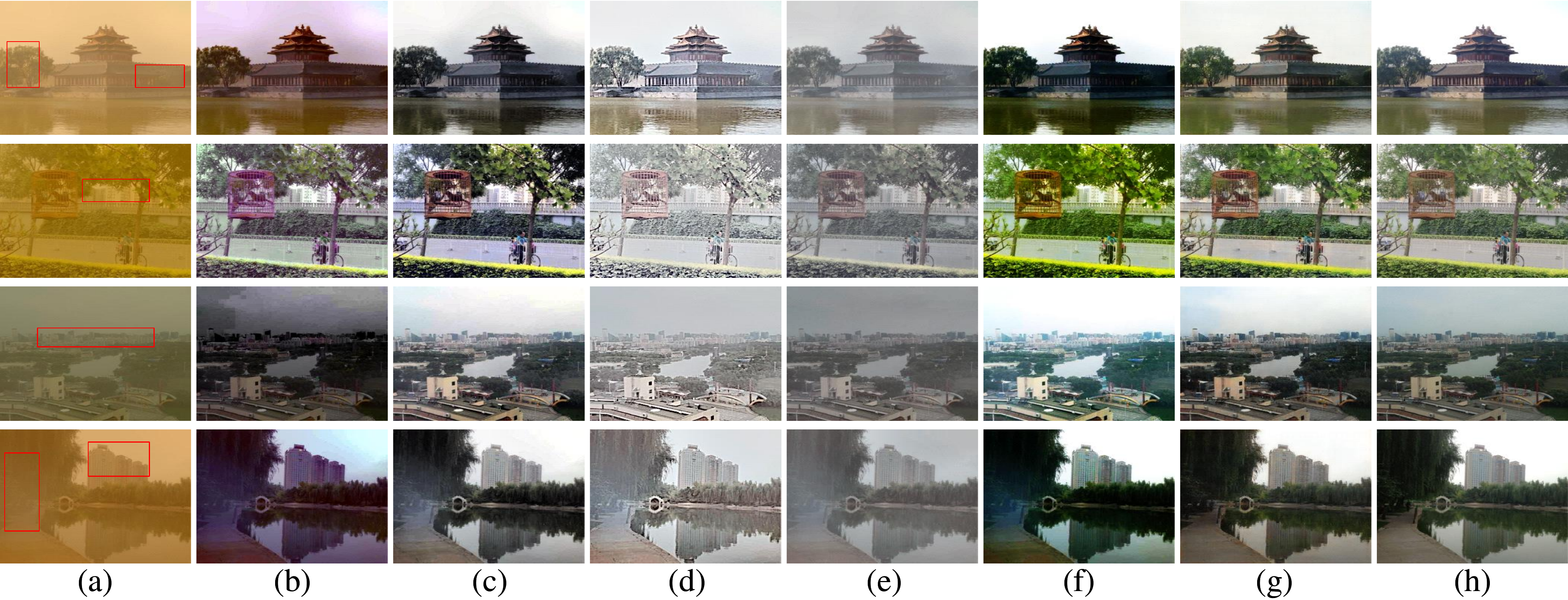}
 \caption{The comparisons on synthetic sand dust images. (a) The sand dust images; (b) CIDC \cite{ref2}; (c) FBE \cite{ref14}; (d) HRDCP \cite{ref4}; (e) NGT \cite{ref10}; (f) TTFIO \cite{ref58}; (g) Pix2pix \cite{ref19}; (h) The reference images.}
 \label{Fig.10} 
\end{figure*}

\begin{figure*}
 \centering
 \includegraphics[width=2\columnwidth]{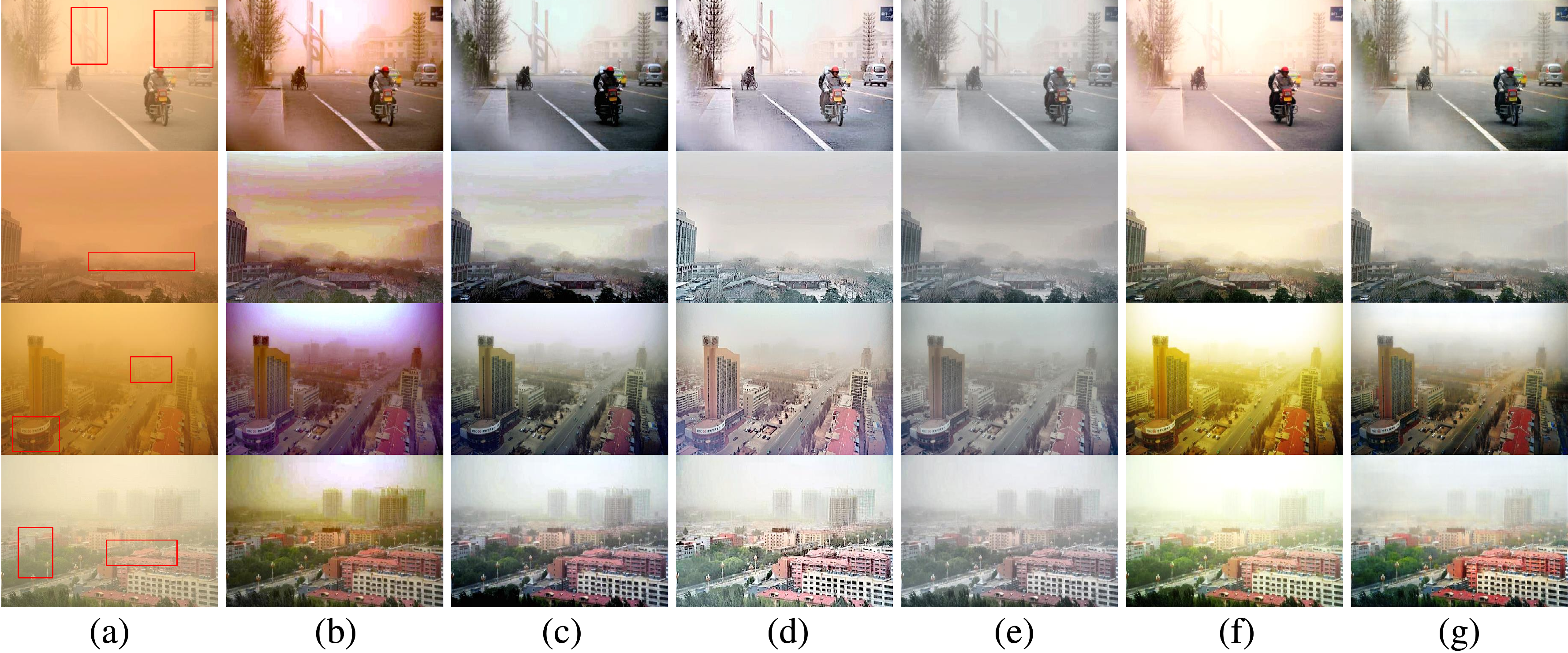}
 \caption{The comparisons on real-world sand dust images from RSTS. (a) The real-world sand dust images; (b) CIDC \cite{ref2}; (c) FBE \cite{ref14}; (d) HRDCP \cite{ref4}; (e) NGT \cite{ref10}; (f) TTFIO \cite{ref58}; (g) Pix2pix \cite{ref19}.}
 \label{Fig.11} 
\end{figure*}

As shown in Figures \ref{Fig.10} and \ref{Fig.11}, we present some visual comparisons on SIRB. Figure \ref{Fig.10} is the comparisons on SIRB-E, and Figure \ref{Fig.11} is the results on RSTS. In Figure \ref{Fig.10}, the results of Pix2pix trained on SIRB-T are visually closer to the reference images. The distribution of real-world sand dust images is complex and changeable, and diversity synthesis strategy is useful for for training deep neural networks with better generalizability. As shown in Figure \ref{Fig.11}, one can see that traditional algorithms generally exist the issues such as color distortion and incomplete dust removal, etc. The approach based on deep learning is better than the existing sand dust removal algorithms in terms of color correction and dust elimination both synthetic images and real-world sand dust images, which shed light the future research of sand dust image reconstruction.

\bibliographystyle{ACM-Reference-Format}
\bibliography{sample-base}
\end{document}